\documentstyle [12pt]{article}
\def\be{\begin{equation}}
\def\ee{\end{equation}}
\def\bea{\begin{eqnarray}}
\def\eea{\end{eqnarray}}
\def\pd{\partial}

\begin{document}

\begin{titlepage}

\title{The General Solution of the Complex Monge-Amp\`ere  Equation
         in a space of arbitrary dimension}

\author{D.B. Fairlie\\
{\it Department of Mathematical Sciences}\\
{\it University of Durham, Durham DH1 3LE}\\
and A.N. Leznov\\
{\it  Institute for High Energy Physics, 142284 Protvino,}\\{\it Moscow Region,
Russia}
{\it and}\\ 
{\it  Bogoliubov Laboratory of Theoretical Physics, JINR,}\\
{\it 141980 Dubna, Moscow Region, Russia}}
\maketitle

\begin{abstract}

A general solution to the Complex Monge-Amp\`ere equation in a space of 
arbitrary dimensions is constructed.

\end{abstract}

\end{titlepage}

\section{Introduction}

 The Complex Monge-Amp\`ere equation in $n$-dimensional space takes the form:
\be 
\det\left|\begin{array}{ccc}
\frac{\pd^2\phi}{\pd y_1\pd\bar y_1} & \ldots& \frac{\pd^2\phi}{\pd y_1\pd
\bar y_n}\\
\vdots & \ddots & \vdots\\
\frac{\pd^2\phi}{\pd y_n\pd\bar y_1} & \ldots& \frac{\pd^2\phi}{\pd y_n\
pd\bar y_n}\\
\end{array}\right|\,=\,0.\label{batman}
\ee
Its real form, which arises from (\ref{batman}) under the assumption that the 
solution
depends only upon $n$ arguments $x_i=y_i+\bar y_i$ was found before by different 
methods \cite{renat},\cite{dbf}. But to the best of our knowledge the general 
solution
of the complex M-A equation (\ref{batman}) is still wanting. The aim of the 
present paper is to fill this gap by using the method of our paper \cite{dbf}
(the solution of real M-A equation in a space of arbitrary dimension) to
obtain and present the general exact solution of the complex version of this 
equation in implicit form. We assume that all functions which we introduce are 
twice differentiable.

\section{Equivalent First Order Equations}

The complex M-A equation, (\ref{batman}) is the eliminant of $n+n$ linear
equations which express the  linear dependence between rows or columns of
the determinantal matrix. They may be written as:
\be 
\sum^n_{i=1} \alpha^i \phi_{y_i,\bar y_k}=0,\quad 
\sum^n_{i=1} \beta^i \phi_{\bar y_i,y_k}=0\label{1}
\ee
where $\phi_{y_i}$ denotes $\displaystyle{\frac{\pd\phi}{\pd y_i}}$ etc.

The next few  rows contain an obvious transformation (\ref{1}) using
only the rules of differentiation together with some new definitions:
\bea
(\sum_{i=1}^n \alpha^i \phi_{y_i})_{\bar y_k}&=
\sum_{i=1}^n \alpha^i_{\bar y_k} \phi_{y_i}\label{R}\\
(\sum_{i=1}^n \beta^i \phi_{\bar y_i})_{y_k}&=&
\sum_{i=1}^n \beta^i_{y_k} \phi_{\bar y_i}\label{R1}
\eea
Introducing two new functions
$$
R=\sum_{i=1}^n \alpha^i \phi_{y_i},\quad 
\bar R=\sum_{i=1}^n \beta^i \phi_{\bar y_i}
$$
considering them as a functions of arguments $R=R(\alpha,y),\ \bar R=\bar R
(\beta,\bar y)$ (under the assumption that $\det J(\alpha,\bar y)$ and $\det 
J(\beta,y)$ are different from zero), we rewrite the equations (\ref{R}) and 
(\ref{R1}) in a equivalent form:
\be
R_{\alpha_i}=\phi_{y_i},\quad \bar R_{\beta_i}=\phi_{\bar y_i}\label{RR}.
\ee
Multiplying each equation (\ref{RR}) respectively by $\alpha_i,\beta_i$,
summing the results and recalling the definitions of the functions, 
$R$, $\bar R$
we come to the conclusion that they 
are homogeneous functions of degree one with respect to the arguments 
$\alpha,\beta$. Introducing the notation
$$
{\alpha_{\pi}\over \alpha_n}=u^{\pi},\quad {\alpha_{\pi}\over \alpha_n}=v^{\pi}
$$
with the convention that Greek indices take values from 1 to $n-1$
we can represent the dependence of the functions $R,\bar R$  in the following 
form:
$$
R=\alpha_n R(u,y),\quad \bar R=\beta_n \bar R(v,\bar y) 
$$
Substituting these expressions into equations (\ref{RR}) we arrive at the
following relations which form the basis of our further investigations:
\bea
\phi_{y_{\beta}}&=&R_{u^{\beta}},\quad \phi_{y_n}=R-\sum u^{\beta}R_{u^{\beta}}
\label{ME1}\\
\phi_{y_{\beta}}&=&\bar R_{v^{\beta}},\quad \phi_{\bar y_n}=\bar R-
\sum v^{\beta}\bar R_{v^{\beta}}
\label{ME2}
\eea

\section{Conditions of selfconsistency (part I)}

Using the condition of equivalence of second mixed derivatives taken in 
different orders, we will be able to disentangle the main system (\ref{ME1}),
(\ref{ME2}) and extract from it a very important system of equations connecting
the functions $u,v$ only. For this goal let us calculate  the second
mixed derivatives of the following pairs of variables $(y_{\alpha},\bar y_
{\beta}),(y_{\beta},\bar y_n),(y_{\bar \beta},y_n),(y_n,\bar y_n)$ and equate 
them.

We have in consequence for the pair $(\bar y_{\beta},y_{\alpha})$:
$$
(R_{u^{\alpha}})_{\bar y_{\beta}}=(\bar R_{v^{\beta}})_{y_{\alpha}},\quad
\sum R_{u^{\alpha},u^{\theta}}u^{\theta}_{\bar y_{\beta}}=\sum 
\bar R_{v^{\beta},v^{\theta}}v^{\theta}_{y_{\alpha}};
$$
for the pair $(y_{\beta},\bar y_n)$:
$$
(R_{u^{\beta}})_{\bar y_n}=(\bar R-\sum v^{\phi}\bar R_{v^{\phi}})_{y_
{\beta}},\quad \sum R_{u^{\beta},u^{\theta}}u^{\theta}_{\bar y_n}=-  
\sum v^{\pi}\sum \bar R_{v^{\pi},v^{\theta}}v^{\theta}_{y_{\beta}};
$$
for the pair $(y_{\bar \beta},y_n)$:
$$
\sum \bar R_{v^{\beta},v^{\theta}}v^{\theta}_{y_n}=-  
\sum u^{\pi}\sum \bar R_{u^{\pi},u^{\theta}}u^{\theta}_{\bar y_{\beta}};
$$
and finally the pair $(y_n,\bar y_n)$ leads to the equations
$$
\sum u^{\beta} \sum R_{u^{\beta},u^{\theta}}u^{\theta}_{\bar y_n}=
\sum v^{\beta}\sum \bar R_{v^{\beta},v^{\theta}}v^{\theta}_{y_n}
$$
Multiplying the first equations respectively by $u^{\alpha},v^{\beta}$,summing
the results and comparing with respectively the second and the third systems
we arrive at  the following systems of equations:
$$
\sum R_{u^{\beta},u^{\theta}}(u^{\theta}_{\bar y_n}+\sum v^{\pi}u^{\theta}_
{\bar y_{\pi}})=0,\quad \sum \bar R_{v^{\beta},v^{\theta}}(v^{\theta}_{y_n}+
\sum u^{\pi}v^{\theta}_{y_{\pi}})=0.
$$
Assuming that $R$ and $\bar R$ are not solutions of the  M-A equations in $(n-1)$
$u,v$ spaces respectively (this case of degeneracy demands additional 
consideration) we conclude that the functions
$(u,v)$  satisfy the following separate system of equations:
\be
u^{\theta}_{\bar y_n}+\sum v^{\pi}u^{\theta}_{\bar y_{\pi}}=0,\quad  
v^{\theta}_{y_n}+\sum u^{\pi}v^{\theta}_{y_{\pi}}=0\label{FUV}
\ee
The system (\ref{FUV}) was solved before \cite{MANII} but for the 
convenience of the reader the next two sections will be devoted to its 
consideration.

The last comment is the following; the hydrodynamic system (\ref{FUV})
is the result of only $2(n-1)$ equations of second mixed derivatives. Namely
combinations of the first, second and third systems. It is not difficult to
check that the equation for the pair $(y_n,\bar y_n)$  is automatically
satisfied.

The $(n-1)^2$ equations remaining unsolved connecting the pairs with barred and 
unbarred Greek indices will be considered in the section 6.
 
\section{The system of hydrodynamic type }

We understand by a system of hydrodynamic type the system (\ref{FUV})  
rewritten below:
\be
v^{\nu}_{y_n}+\sum u^{\mu} v^{\nu}_{y_{\mu}}=0,\quad 
u^{\mu}_{\bar y_n}+\sum v^{\nu} u^{\mu}_{\bar y_{\nu}}=0\label{I}
\ee

Two propositions with respect to this system will be crucial in what
follows.

Proposition 1.
The pair of operators:
\be
D=\frac{\partial}{\partial y_n}+\sum u^{\mu} \frac{\partial}{\partial y_{\mu}},
\quad
\bar D=\frac{\partial}{\partial \bar y_n}+\sum v^{\nu} \frac{\partial}{\partial 
\bar y_{\nu}}\label{II}
\ee
are mutually commutative if $(u^{\mu},v^{\nu})$ are solutions of the system 
(\ref{I}).

Acting with the help of operators $(D,\bar D)$ on the second and the first 
equations of (\ref{I}) respectively we come to conclusion that $2(n-1)$
functions:
\be
\bar D(v^{\nu})=v^{\nu}_{\bar y_n}+\sum v^{\mu} v^{\nu}_{\bar y_{\mu}},\quad 
D(u^{\mu})=u^{\mu}_{y_n}+\sum u^{\nu} u^{\mu}_{y_{\nu}}
\ee
are also solutions of the first and the second system of equations (\ref{I}).

As a corollary we obtain the following  

Proposition 2
\be
v^{\nu}_{\bar y_n}+\sum v^{\mu} v^{\nu}_{y_{\bar \mu}}=V^{\nu}(v;\bar y),\quad 
u^{\mu}_{y_n}+\sum u^{\nu} u^{\mu}_{y_{\nu}}=U^{\mu}(u;y)\label{BA}
\ee
Indeed the $n$ sets of variables $(1,u)$, and $(1,v)$ respectively satisfy  a 
linear system of algebraic equations of $n$ equations, the matrix of which 
coincides with the Jacobian matrix
$$
J=\det_n \left|\begin{array}{cccc} v^1 & \dots & v^{n-1} & V^{\nu} \\
                     y_1 & \dots & y_{n-1} & y_n \end{array}\right|
$$
which in the case of a non-zero solution of the linear system must
vanish. So Proposition 2 is proved.

Compared with (\ref{I}) (\ref{BA}) is  an inhomogeneous system of hydrodynamic
equations separated into functions  $(u,v)$.

With respect to the generators $D,\bar D$ all functions of $2n$ dimensional space
may be divided into the following subclasses: functions of  general position
$F,\  D F\neq 0,\ \bar D F\neq 0$, the "holomorphic" functions $f,\ \bar D f=0,
\ D f
\neq 0$, "antiholomorphic" ones $\bar f, D \bar f=0,\bar d \bar f\neq 0$ and 
"central" functions $f^0$ which are holomorphic and antiholomorphic 
simultaneously ;$\bar D f^0=D f^0=0$. 
Each central function may be represented in the form:
\be
f^0\,=\,f^0(Q)\,=\,f^0(P)\,=\,g^0(\psi)\label{III}
\ee
The  reader will find the definitions of the  functions $Q,\ P,\ \psi$  in the next section.

\section{General solution of the hydrodynamic system}

Suppose we have the following system of equations defining implicitly $(n-1)$
 unknown 
functions $(\psi)$ in $(2n)$ dimensional space co-ordinatized by $(y,\bar y)$:
\be 
Q^{\nu}(\psi;y)=P^{\nu}(\psi;\bar y)\label{D}
\ee
The number of equations in (\ref{D}) coincides with the number of unknown 
functions $\psi^{\alpha}$.

With the help of the usual rules of differentiation of  implicit functions
we find from (\ref{D}):
\bea
\psi_y=(P_{\psi}-Q_{\psi})^{-1} Q_y,\quad \psi_{\bar y}=-(P_{\psi}-Q_{\psi})^
{-1}P_{\bar y}\label{DD}
\eea
Let us assume, that between $n$ derivatives with respect to barred and unbarred
variables there exists the linear dependence:
\be
\sum^n_1 c_i \psi^{\alpha}_{y_i}=0,\quad \sum^n_1 d_i \psi^{\alpha}_{\bar y_i}=0
\label{LC}
\ee
and analyse the consequences following from these facts.

Assuming that $c_n\neq 0,d_n\neq 0$, dividing each equation of the left and 
right 
systems respectively by them and introducing the notations $u^{\alpha}=
{c_{\alpha}\over c_n},v^{\alpha}={d_{\alpha}\over d_n}$ we rewrite the last 
systems in the form:
\be
\psi^{\alpha}_{y_n}+\sum^{n-1}_1 u^{\nu} \psi^{\alpha}_{y_{\nu}}=0,\quad
\psi^{\alpha}_{\bar y_n}+\sum^{n-1}_1 v^{\nu} \psi^{\alpha}_{\bar y_{\nu}}=0
\label{MS} 
\ee
Substituting the values of the derivatives from (\ref{DD}) and multiplying the 
result by the matrix $(P_{\phi}-Q_{\phi})$ on the left we obtain:
\be
Q^{\alpha}_{y_n}+\sum^{n-1}_1 u^{\nu} Q^{\alpha}_{y_{\nu}}=0,\quad
P^{\alpha}_{\bar y_n}+\sum^{n-1}_1 v^{\nu} P^{\alpha}_{\bar y_{\nu}}=0
\label{D1}
\ee
From the last equations it immediately follows:
\be
u^{\nu}=-(Q_y)^{-1} Q_{y_n},\quad v^{\nu}=-(P_{\bar y})^{-1} P_{\bar y_n}
\label{UV}
\ee
We see that if we augment the initial system (\ref{D}), by $(n-1)$ vector 
functions
$(u,v)$ defined by (\ref{UV}) then operators of differentiation $D,\bar D$
defined by (\ref{II}) in connection with (\ref{MS})  annihilate each 
$\psi$ either as a $P$ or a $Q$ function:
\be
D \psi=\bar D \psi=D Q=D P=\bar D Q=\bar D P=0 \label{VIC}
\ee
These  equations (\ref{VIC}) explain the notations in the last formula 
(\ref{III}) of the previous section

This means that $D\bar f(\phi,\bar y)=\bar D f(\phi, y)=0$. As a direct
corollary of this fact $Dv=\bar D u=0$, so the generators $D,\bar D$
constructed above are mutually commutative. Thus we have the found general 
solution of the hydrodynamic system.

\section{Conditions of selfconsistency (part II)}

The general solution of the hydrodynamic system depends upon $2(n-1)$ 
arbitrary functions $(P,Q)$ each dependent upon $2n-1$ independent arguments. 
This collection of arbitrary functions is more much (exept for the case $n=2$) 
than  is necessary for the general solution of the M-A equation. So it is
possible to expect that other conditions of selfconsistency (unused up to now) 
reduce it up to two functions each of $2n-1$ indepent arguments.

We begin from the remaining unsolved $(n-1)^2$ equations of section 3, 
rewritten below:
\be
\sum R_{u^{\alpha},u^{\theta}}u^{\theta}_{\bar y_{\beta}}=\sum 
\bar R_{v^{\beta},v^{\theta}}v^{\theta}_{y_{\alpha}}\label{gbg}
\ee

For this and all calculations below  knowledge of the explicit expressions 
for the derivatives of the functions $(u,v)$ functions will be necessary.
We have in consequence:
$$
u_{y_{\alpha}}=-Q_y^{-1}(Q_{y_n,y_{\alpha}}+\sum Q_{y_n,\psi^{\beta}}\psi^
{\beta}_{y_\alpha}-Q_{y,y_{\alpha}}Q_y^{-1}Q_{y_n}-\sum Q_{y,\psi^{\beta}}\psi^
{\beta}_{y_\alpha}Q_y^{-1}Q_{y_n})\equiv
$$
$$
-Q_y^{-1}(D Q_{y_{\alpha}})+Q_y^{-1}(D Q_{\psi})(P_{\psi}-Q_{\psi})^{-1}
Q_{y_{\alpha}}
$$
$$
v_{\bar y_{\alpha}}=-P_{\bar y}^{-1}(\bar D P_{\bar y_{\alpha}})+P_y^{-1}
(\bar D P_{\psi})(P_{\psi}-Q_{\psi})^{-1}P_{\bar y_{\alpha}} 
$$
By the same technique we calculate $u_{\bar y},v_y$ with the result:
$$
u_{\bar y}=-Q_y^{-1}(D Q_{\psi})(P_{\psi}-Q_{\psi})^{-1}P_{\bar y},\quad
v_y=-P_{\bar y}^{-1}(D P_{\psi})(P_{\psi}-Q_{\psi})^{-1}Q_y
$$
Substituting the calculated values of derivatives into (\ref{gbg}), we obtain 
in matrix notation:
$$
R_{u^{\alpha},u}Q_y^{-1}(D Q_{\psi})(P_{\psi}-Q_{\psi})^{-1}P_{\bar y_{\beta}}=
\bar R_{v^{\beta},v}P_{\bar y}^{-1}(D P_{\psi})(P_{\psi}-Q_{\psi})^{-1}Q_{y_
{\alpha}}
$$
or after moving the matrices $P_{\bar y_{\beta}},Q_{y_{\alpha}}$ to the left 
and right respectively we obtain the matrix equation:
\be
(Q^T_y)^{-1}R_{u,u}Q_y^{-1}(D Q_{\psi})(P_{\psi}-Q_{\psi})^{-1}= 
[(P^T_{\bar y})^{-1}\bar R_{v,v}P_{\bar y}^{-1}(D P_{\psi})(P_{\psi}-Q_{\psi})^
{-1}]^T\label{GBG}
\ee
where $T$ denotes transpose.

Now let us consider results which follow from  equating  the second mixed 
derivatives with unbarred Greek indices.  Calculations similar to the previous 
lead to the final result containing ${(n-1)(n-2)\over 2}$ equations:
$$
-(Q^T_y)^{-1}R_{u,u}Q_y^{-1}(D Q_y) Q_y^{-1}+(Q^T_y)^{-1}R_{u,u}Q_y^{-1}
(D Q_{\psi})(P_{\psi}-Q_{\psi})^{-1}+
$$
\be
(Q^T_y)^{-1}R_{u,y}Q_y^{-1}=[ ...]^T \label{GG}
\ee
where by dots in the quadratic brackets of the right side we denote the matrix
 of the left 
hand side of the last equation.

The same calculations for mixed partial derivatives with barred Greek indices leads to 
result:
$$
-(P^T_{\bar y})^{-1}\bar R_{v,v}P_{\bar y}^{-1}(\bar D P_{\bar y}) P_{\bar y}^
{-1}+(P^T_{\bar y})^{-1}\bar R_{v,v}P_{\bar y}^{-1}
(\bar D P_{\psi})(P_{\psi}-Q_{\psi})^{-1}+
$$
\be
(P^T_{\bar y})^{-1}\bar R_{v,\bar y}P_{\bar y}^{-1}=[ ...]^T\label{BGBG}
\ee

Summing the systems (\ref{GG}) and (\ref{BGBG}) and taking into account
(\ref{GBG}) we eliminate terms with the factors $(P_{\psi}-Q_{\psi})^{-1}$ and 
come to the equation
\bea
-(Q^T_y)^{-1}R_{u,u}Q_y^{-1}(D Q_y) Q_y^{-1}+(Q^T_y)^{-1}R_{u,y}Q_y^{-1}-[ ...]
^T=\nonumber\\
-(P^T_{\bar y})^{-1}\bar R_{v,v}P_{\bar y}^{-1}(\bar D P_{\bar y}) P_{\bar y}
^{-1}+(P^T_{\bar y})^{-1}\bar R_{v,\bar y}P_{\bar y}^{-1}-[ ...]^T\equiv A^0
\eea

Indeed the left hand side of the last equality depends upon the arguments 
$(u,y)$, while the right hand side depends upon the arguments $(v,\bar y)$ and
in view of the comments in the end of the section 4 the antisymmetrical 
$(n-1)\times (n-1)$ matrix $A^0$ is a central function.

The combination of the equations (\ref{GBG}) and (\ref{GG}) leads to:
\be
(Q^T_y)^{-1}R_{u,u}Q_y^{-1}(D Q_{\psi})(P_{\psi}-Q_{\psi})^{-1}-  
(P^T_{\bar y})^{-1}\bar R_{v,v}P_{\bar y}^{-1}(\bar D P_{\psi}
(P_{\psi}-Q_{\psi})^{-1}=A^0\nonumber
\ee
After multiplication of last equation by the matrix $(P_{\psi}-Q_{\psi})$ on
the right and the observation that the left hand side of the equation arising is
the difference of holomorphic and antiholomorphic functions we solve it with
the final result:
\be 
(Q^T_y)^{-1}R_{u,u}Q_y^{-1}(D Q_{\psi})=A^0Q_{\psi},\quad
(P^T_{\bar y})^{-1}\bar R_{v,v}P_{\bar y}^{-1}(\bar D P_{\psi})=A^0P_{\psi}
\label{BUIT2}
\ee
This last system together with equation defining $A^0$ is now the subject of  
further investigation.

\section{Solution of selfconsistency equations}

At first we consider in detail the simplest examples $n=2$ solved before 
\cite{ma2}, which will serve as a guess for a form of solution in the general 
case  of arbitrary $n$.

\subsection{The case n=2}

In this case Greek index takes only one value $1$, the antisymmetrical matrix
$A^0$ is equal to zero. In this sense it escapes from from the general case. In 
spite
of this the calculations in this case make a good exercise useful for further 
consideration.
In this case the system of equations (\ref{ME1}), (\ref{ME2}) takes the 
following form:
$$
\phi_{y_1}=R_u,\quad \phi_{y_2}=R-uR_u,\quad \phi_{\bar y_1}=\bar R_v,\quad 
\phi_{\bar y_2}=\bar R-v\bar R_v
$$
The second mixed partial derivatives of the pairs $(y_1,\bar y_2),(\bar y_1,y_2)$,
and $(\bar y_2,y_2)$ have as their corollary the hydrodynamical system of 
equations:
$$
v_{\bar y_2}+uv_{\bar y_1}=0,\quad v_{y_2}+vu_{y_1}=0
$$
as in the general case; the general solution of which is given in connection 
with the results of section 5 by the formulae {\ref{UV}):
$$
u=-{Q_{y_2}\over Q_{y_1}},\quad v=-{P_{\bar y_2}\over P_{\bar y_1}}
$$

There is only one equation of selfconsistency, cannecting the barred and 
unbarred index $1$:
\be
(R_u)_{\bar y_1}=(\bar R_v)_{y_1},\quad R_{uu}u_{\bar y_1}=\bar R_{vv}v_{y_1},
\quad R_{uu}u_{\psi}\psi_{\bar y_1}=\bar R_{vv}v_{\psi}\psi_{y_1}\label{F4}
\ee

Substituting into (\ref{F4}) the known values of the derivatives of the 
function $\psi$  (\ref{DD}) we pass to the final equation of interest:
$$
{R_{uu}u_{\psi}\over Q_{y_1}}=-{\bar R_{vv}v_{\psi}\over P_{\bar y_1}}=
A^0_{\psi}
$$
Indeed the left hand side of the last equality is a holomorphic function, 
the right hand side antiholomorphic one. Thus $A^0_{\psi}$ is a central 
function.

The equations of selfconsistenct for the pairs $(y_1,y_2),(\bar y_1,\bar y_2)$
may be manipulated to the following attractive form:
\be
D R_u=R_{y_1},\quad \bar D \bar R_v=R_{\bar y_1}\label{F2}
\ee
Considering now $R_u=R_u(\psi;y_1,y_2)$ and $\bar R_v=\bar R_v(\psi;\bar y_1,
\bar y_2)$ we resolve equations containing $A^0_{\psi}$ function in the form:
\bea
R_u&=&\Theta_{y_1}(A;y_1,y_2),\quad Q=\Theta_A(A;y_1,y_2)\nonumber\\
R_v&=&\bar \Theta_{\bar y_1}(A;\bar y_1,\bar y_2),\quad P=-\bar \Theta_A(A;
\bar y_1,\bar y_2)\label{RT}
\eea

It remains only to check the equalities (\ref{F2}). Let us distinguish by the
upper indices $u,A$ the derivatives
$\frac{\partial^u}{\partial y_i},\frac{\partial^A}{\partial y_i}$ corresponding 
to the partial derivatives of the space coordinates $(y,\bar y)$ keeping
 $u,(v)$ constant in first case and $A$ at a constant value in the second. The 
 equality which has to be checked in this notations is
$$
\frac{\partial^u}{\partial y_1} R_u=\frac{\partial}{\partial u}
\frac{\partial^u}{\partial y_1} R\nonumber
$$
Keeping in mind that $D^u A=0$ ($A$ is a central function) and the definition of
all values involved in terms of the  function $\Theta$, we obtain in consequence for 
the right hand side of the last equality;
\bea
(D^u R_u)_u&=&(D^u \frac{\partial^A}{\partial y_1}\Theta)_u=
(D^u \frac{\partial^A}{\partial y_1}\Theta)_A A_u=\nonumber\\
(\frac{\partial^{2A}}{\partial y_1\partial y_2}&+&u\frac{\partial^{2A}}
{\partial y_1\partial y_1})\Theta)_A A_u=\Theta_{y_1,y_1}+(\Theta_{A,y_1,y_2}-
{\Theta_{A,y_2}\over \Theta_{A,y_1}}\Theta_{A,y_1,y_1}) A_u=\nonumber\\
\Theta_{y_1,y_1}&+&\Theta_{A,y_1} A_u \frac{\partial^Au}{\partial y_1}\nonumber
\eea
In all transformations above we have not written the upper index $A$ with
 respect
to derivatives of the space coordinates $y$ applied to the function $\Theta$ .

Similar calculations for the left hand side leads to:
$$
\frac{\partial^u}{\partial y_1} \Theta_{y_1}=\Theta_{y_1,y_1}+\Theta_{A,y_1}A_u 
\frac{\partial^Au}{\partial y_1}  
$$
which shows that equalities (\ref{F2}) are satisfied.

But (\ref{F2}) in its turn is an equation of second order with respect to the 
unknown  function $R$. We rewrite it in explicit form substituting instead of 
$R_u$ its value from (\ref{RT}):
\bea
(\frac{\partial^{2}}{\partial y_1\partial y_2}&+&u\frac{\partial^{2}}
{\partial y_1\partial y_1})\Theta=R^u_{y_1}\equiv R_{y_1}+R_A A_{y_1}=\nonumber
\\
R_{y_1}-{u_{y_1}\over u_A}R_A&=&R_{y_1}-u_{y_1}R_u=R_{y_1}-u_{y_1}\Theta_{y_1}
\eea
In the process of the evaluation of the last expression the crucial element was
the calculation
of $A_{y_1}$ maintaining the value of $u$ fixed . It was achieved by direct 
differentiation of the definition of $u$ rewritten in the form:
$$
\Theta_{A,y_2}+u\Theta_{A,y_1}=0
$$
with respect to the argument $y_1$ (under  fixed $u$) and regrouping of the 
terms arising.

Preserving in the last equality the first and the last terms we arrive at the 
equation for the function $R$  in integrable form. The result of its 
integration
determines the function $R$ in terms of the function $\Theta$ in a very 
attractive form:
$$
R=D \Theta,\quad \bar R=\bar D \bar \Theta
$$
Substituting these expressions in the equations connected derivatives of the 
solution 
of the M-A equation (\ref{ME1}), (\ref{ME2}) with $R,\bar R$ functions we
obtain finally
$$ 
\phi_{y_1}=\Theta_{y_1},\quad \phi_{y_2}=\Theta_{y_2},\quad \phi_{\bar y_1}=
\bar \Theta_{\bar y_1},\quad \phi_{\bar y_2}=\bar \Theta_{\bar y_2}
$$

Theorem:

Let the function $A$ be determined implicitly by the equation:
$$
\Theta_A(A;y_1,y_2)=-\bar \Theta_A(A;\bar y_1,\bar y_2)  
$$
where $\Theta,\bar \Theta$ are arbitrary functions of their 3 arguments,
then selfconsistent derivatives of the function $\phi$ satisfying the complex 
M-A equation in two dimension are determined with the help of the formulae
$$
\phi_{y_1}=\Theta_{y_1},\quad \phi_{y_2}=\Theta_{y_2},\quad \phi_{\bar y_1}=
\bar \Theta_{\bar y_1},\quad \phi_{\bar y_2}=\bar \Theta_{\bar y_2}
$$
 
\section{General case of arbitrary $n$}

We will not present the calculations which are not very simple  allowing 
us to obtain the main 
result for arbitrary $n$ but give its formulation and prove it directly. This 
turns out to be much  easier.

Theorem:

Let the the set of the functions $\psi^{\alpha}(y;\bar y)$ be determined 
implicitly by the following set of equations, the number of which coincides 
with the number of $\psi$ functions:
$$
\Theta_{\psi^{\beta}}(\psi;y)=-\bar \Theta_{\psi^{\beta}}(\psi;\bar y)
$$
where $\Theta,\bar \Theta$ are arbitrary functions of their $(2n-1)$ arguments.
Then selfconsistent derivatives of the function $\phi$ satisfying the complex 
M-A equation in $n$ dimension are determined with the help of the formulae:
$$
\phi_y=\Theta_y,\quad \phi_{\bar y}=\bar \Theta_{\bar y}
$$

Let us  first check the conditions of selfconsistency of the second mixed
partial derivatives with the same (barred, unbarred) indices. We have consequently
($(y_1,y_2)$ - arbitrary two coordinates):
$$
(\phi_{y_1})_{y_2}=\Theta_{y_1,y_2}+\sum \Theta_{y_1,\psi^{\nu}} \psi^{\nu}_{y_2}
=\Theta_{y_1,y_2}+\sum \Theta_{y_1,\psi}(\Theta_{\psi,\psi}+\bar \Theta_{\psi,
\psi})^{-1} \Theta_{y_2,\psi}
$$
The matrix $(\Theta_{\psi,\psi}+\bar \Theta_{\psi,\psi})^{-1}$ is obviously 
symmetric  so the last expression is symmetric with respect to a
permutation of the indices $(1,2)$ . Thus second mixed partial derivatives with the same kind of indices are self consistent.

Now let us calculate the mixed derivatives with indices of  different kinds:
$$
(\phi_{y_i})_{\bar y_k}=\sum \Theta_{y_i,\psi^{\nu}} \psi^{\nu}_{\bar y_k}
=\sum \Theta_{y_i,\psi}(\Theta_{\psi,\psi}+\bar \Theta_{\psi,\psi})^{-1} 
\bar \Theta_{\bar y_k,\psi}
$$
The result of the calculation in the opposite order gives exactly the same 
result
also as corollary of symmetry of the same matrix.
 
Finally let us multiply the second mixed partial derivatives calculated above 
by
$\beta^k$ and sum the result. We see that from right hand side the term 
$\bar D (\bar \Theta_{\psi}$, which is equal to zero
always arises.

Thus we have proved that between the the rows  ( and columns) of the 
determinantal 
matrix  linear dependence occurs and the so equation of M-A in $n$ dimensional
space is satisfied.

\section{Outlook}

The main result of the present paper is in the Thorem of the previous section,
giving the possibility of finding a general solution of the homogeneous Complex 
M-A
equation (\ref{batman}) in implicit form. We specially emphasize that we can't
say that we have found all solutions of this equation but only those in which
the number of arbitrary functions and their functional dependence are sufficient
for
the statement of the problem of the solution of the M-A equation in terms of initial data in
the sense of Cauchy-Kovalevski. This solution is the most nondegenerate and
excludes solutions of the shock wave type. Moreover only after detailed analysis of all 
the many
assumptions made  and the  precise consideration
of the corollaries which follow from the results of section 6 (which we have 
ommitted
in this paper) it will be possible to clarify the situation and increase our 
understanding of what collection of solutions is contained in the construction 
of the present paper.

No less important and interesting is the general solution of the hydrodynamic 
(\ref{FUV}) system of equations. It turns out that, using it as a basis, it is
possible to generalise two-dimensional theory of integrable systems (
based upon representation theory of semisimple algebras) \cite{LAN} in the
multidimensional case \cite{MANII}.
  
Using the hydrodynamic system of equations it is possible to solve the Complex
Bateman (Complex Universal) equation \cite{combat}. (This equation serves as a 
Lagrangian for the complex M-A) As the present solution of the $M-A$ 
equation and that of the  Complex Bateman correspond to  different reductions
of the general solution of hydrodynamic system, we can't 
exclude the possibility that other interesting reductions exist which are 
connected with multidimensional (in the sense of the number of unknown 
functions) systems of equations of M-A and Universal equations hitherto
unknown.

\section*{ Acknowledgements.}

One of the the authors (ANL) is indebted to the Center for Research on 
Engenering and Applied Sciences (UAEM, Morelos, Mexico) for its hospitality and
Russian Foundation of Fundamental Researches (RFFI) GRANT N 98-01-00330 for 
partial support.

\end{document}